# Iterative Distortion Cancellation Algorithms for Single-Sideband Systems

JUN DONG, TIANWAI BO*, ZHUO WANG, HAOLEI GAO, ZHONGWEI TAN, YI DONG

*Key Laboratory of Photonic Information Technology, Ministry of Industry and Information Technology, School of Optics and Photonics, Beijing Institute of Technology, Beijing 10081, China*
*\*tanwai@bit.edu.cn*

**We propose an iterative distortion cancellation algorithm to digitally mitigate the impact of double-sideband dither signal amplitude from the automatic bias control module on Kramers-Kronig receivers without modifying physical layer structures. The algorithm utilizes the KK relation for initial signal decisions and reconstructs the distortion caused by dither signals. Experimental tests in back-to-back showed it improved tolerance to dither amplitudes up to 10% $V_\pi$. For 80-km fiber transmission, the algorithm increased the receiver sensitivity by more than 1 dB, confirming the effectiveness of the proposed distortion cancellation method.**

https:// doi.org/10.0000/OL.00.000000

The rapid development of AI applications has driven the demand for high-capacity fiber-optic communication systems. For short-distance interconnections, direct detection offers a cost-effective solution, which has attracted extensive research interest due to its structural simplicity and cost-effectiveness [1,2]. For high-capacity short-reach transmissions, such as data center interconnects (DCI) and metro networks, single-sideband (SSB) signals with direct detection provide a promising solution, as chromatic dispersion (CD) can be digitally compensated after field recovery. To recover the field, researchers proposed the Kramers–Kronig (KK) receiver to address the signal–signal beating interference (SSBI) impairment that occurs during single-ended photodiode (PD) reception [3]. Additionally, the most common SSB implementation employs an IQ modulator-based optical transmitter, which has a simple structure, high signal quality, and a wide adjustment range of carrier-to-signal power ratio (CSPR) [4]. Typically, IQ modulators utilize dither-based automatic bias control (ABC) techniques to eliminate bias voltage drift resulting from temperature fluctuations, mechanical vibrations, and other environmental changes [5,6]. A unique limitation of the KK receiver is its strict requirement for minimum-phase conditions. This requires the transmitted signal to strictly exhibit SSB features, with the carrier power exceeding that of the information signal [7]. Nonetheless, using double-sideband (DSB) dither signals from the ABC device for system control can violate the minimum-phase condition [8]. Since the ABC module requires sufficient strength of the dither signal (usually no smaller than 2% of the half-wave voltage of the modulator, $V_\pi$), it is of great importance to handle the distortions induced by the dither signals. For the nonlinear distortions, researchers enhanced square-root KK processing to match device real characteristics, thereby reducing the impact of nonlinear impairments [9]. But the general nonlinear compensation method is shown to be inefficient in eliminating this specific distortion [10]. Thus, the challenge remains to develop a more efficient and practical solution for compensating for these distortions while minimizing computational demands.

This letter discusses using the ABC device for stable SSB transmission, noting performance degradation from signal distortion caused by dither signals. It then introduces a low-complexity distortion cancellation algorithm that iteratively removes distortions, recovering signals from the KK receiver. This solution doesn't require hardware changes, instead enhancing system performance with a moderate increase in offline digital signal processing (DSP) complexity.

For SSB signals, the received photocurrent after square-law detection can be expressed as:

$$I = |E_0 + E_s|^2 = |E_0|^2 + |E_s|^2 + 2\text{Re}\{E_s\}E_0 \quad (1)$$

where $E_0$ is the optical carrier, and $E_s$ is the electric field of the optical SSB signal. The KK algorithm enables the reconstruction of the SSB signal under the minimum-phase condition. However, when an ABC system introduces dither signals $m_d \cos(\omega_d t + \theta)$ at the transmitter, where $m_d$, $\omega_d$, and $\theta$ denote the amplitude, angular frequency of the dither signal, and the delay angle between the dither and the transmission signal respectively, the DSB dither signal will violate the minimum-phase condition. Meanwhile, the received photocurrent becomes:

$$\begin{aligned}
I' &= |E_0 + E_s + m_d \cos(\omega_d t + \theta)|^2 \\
&= |E_0|^2 + |E_s|^2 + 2E_0 \text{Re}\{E_s\} \\
&\quad + 2m_d \text{Re}\{E_s\}\cos(\omega_d t + \theta) + 2m_d E_0 \cos(\omega_d t + \theta) \\
&\quad + m_d^2 \cos^2(\omega_d t + \theta)
\end{aligned} \quad (2)$$

Here, the first three terms are the same as the intensity waveform of a perfect optical SSB signal given by Eq. (1). The remaining three terms are the carrier-dither, signal-dither, and dither-dither beat interferences, respectively. Among them, the dither-dither beat signal can be negligible due to the small magnitude of $m_d$, while the carrier-dither and signal-dither beat play critical roles in degrading the quality of the signal. As illustrated in Fig. 1, the signal-dither beat interference generates in-band distortions. Fortunately, the distortions can be estimated from the signal as well as the dither frequency as they are copies of the signal, $\text{Re}\{E_s\}$, frequency-shifted by $-f_d$ and $f_d$, respectively. As we can estimate the distortive terms and subtract them from the received photocurrent, the signal quality can be enhanced significantly.

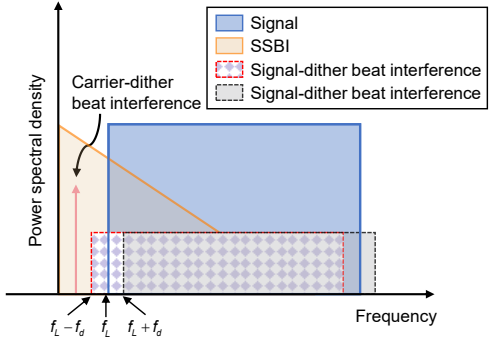

Fig. 1. The spectral components of the photocurrent, $I'$, as defined in Eq. (2). $f_L$ is the lowest frequency of the signal, $f_d$ is the dither frequency.

For this purpose, we utilize an iterative distortion cancellation scheme with the KK algorithm. First, we ignore the existence of the distortive terms and apply the KK algorithm to recover the signal, named by $E_s'$. Here, $E_s' = E_s + \Delta E_s$, where $\Delta E_s$ is the deviation between the estimated and original $E_s$. Then, we reconstruct the distorted signal based on $E_s'$. The distortion can be reconstructed with:

$$I_{distortion} = 2m_d E_0 \cos(\omega_d t + \theta) + 2m_d \text{Re}\{E_s'\}\cos(\omega_d t + \theta) \quad (3)$$

Finally, the distortion is subtracted from the received photocurrent, and more reliable decisions can be made with it.

$$I' - \alpha I_{distortion} = I - 2\alpha m_d' \text{Re}\{\Delta E_s\}\cos(\omega_d t + \theta) \\ + 2(m_d - \alpha m_d')E_0 \cos(\omega_d t + \theta) \\ + 2(m_d - \alpha m_d')\text{Re}\{E_s\}\cos(\omega_d t + \theta) \quad (4)$$

where $\alpha$ is an amplitude scaling factor, we determine the ideal value as $\alpha = m_d/m_d'$. As iteration goes on, we expect $\Delta E_s$ to approach 0 and $\alpha m_d'$ to approach $m_d$. Therefore, $I' - \alpha I_{distortion}$ approaches $I$.

To investigate how this system undermines the minimum-phase condition of the KK receiver, we simulated the transmission of 20-Gbaud orthogonal frequency division multiplexing (OFDM) modulated signals in a back-to-back (BTB) scenario. As shown in Fig. 2, the distribution of various frequency components is clearly visible. In fact, the received signal contains numerous nonlinear frequency components. Among them, the double-sideband dither signals destroy the single-sideband characteristics of the optical signal; the signal-dither beat interference is an in-band distortion covering the entire signal bandwidth; and the carrier-dither beat interference causes significant fluctuations in the signal. These nonlinear effects significantly impact the accuracy of KK relation recovery, rendering the KK algorithm incapable of accurately reconstructing the signal and resulting in severe transmission impairments.

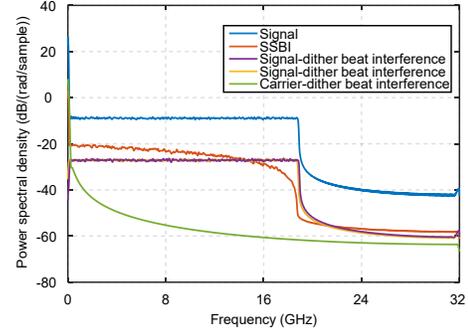

Fig. 2. Simulation of PSD fitting for the received signal.

We propose dither-signal beat interferences cancellation (DSBIC) scheme to address the above issues. The principle of the algorithm is shown in Fig. 3. The compensation algorithm operates in two consecutive stages. First, the KK algorithm is applied to obtain an initial signal decision, and the DC component is extracted to estimate the carrier amplitude of the $E_s$. Subsequently, two dither signals from the ABC module are digitally approximated as $\cos(\omega_I t)$ and $\cos(\omega_Q t)$. In this module, DSBIC is performed to find the optimal amplitude and angle of the dither signals. For each tone, the amplitude and angle are blindly searched in a grid with step sizes of 0.02 rad and 0.52 rad (25 × 12), respectively. The number of preset values and the step sizes are chosen to balance the algorithm's performance with its complexity, which will be discussed below. After the two-dimensional scan, the optimal amplitude and angle for the reconstructed dither are determined by minimizing the bit error rate (BER). We extract the amplitude of the second-order frequency of the dither signal using a cross-correlation method to determine the $m_d$, and from this, we can find the optimal $\alpha$. Finally, the $\alpha$ is used to normalize the reconstructed impairment signal, which is then subtracted from the received photocurrent signal to eliminate the distorted signal.

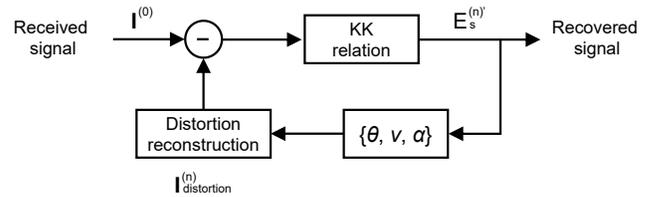

Fig. 3. Signal processing using the proposed compensation algorithm based on the KK relation.

We can easily see that the computational complexity of DSBIC scales with the number of preset values for the amplitude and angle of the tones evaluated in the blind grid search loops. For simplicity, we can use the number of real-valued multiplications as a figure of merit, considering that the complexity of other operations, such as additions, takes a small proportion of the total complexity. Each loop search requires $4n$ real multiplications for testing the distorted signal construction and $n^2$ for error calculation, where $n$ represents the length of the received signal. Note that the test dither wave can be efficiently implemented by using a look-up table. Therefore, the overall computational complexity of the DSBIC algorithm can be estimated as $N_m N_\theta (4n + n^2)$, where factor $N_m$ and $N_\theta$ represent the number of amplitudes and angles considered in the loops, respectively.

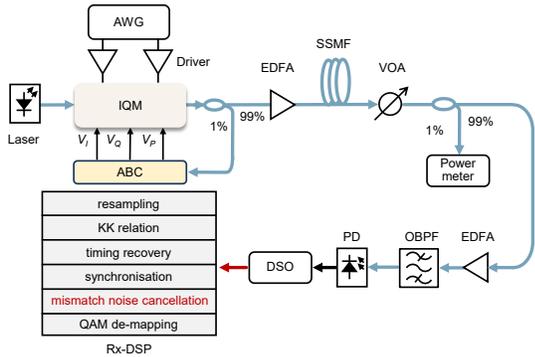

Fig. 4. The experimental setup. AWG: arbitrary waveform generator, IQM: in-phase and quadrature modulator, SSMF: standard single-mode fiber, VOA: variable optical attenuator, EDFA: erbium-doped fiber amplifier, OBPF: optical bandpass filter, PD: photodetector, DSO: digital storage oscilloscope.

We perform transmission experiments to evaluate the effectiveness of the proposed DSBIC technique for optical SSB transmission. Fig. 4 shows the experimental setup. On the transmitter (Tx) side, a random bit sequence is generated offline in MATLAB, then mapped to 16-ary quadrature amplitude modulation (QAM) symbols, and loaded onto an OFDM signal consisting of 316 positive-frequency data points with an FFT size of 1024 subcarriers. The cyclic prefix is 32. Therefore, the total bit rate is approximately 80 Gb/s. The signal is converted to the analog domain using a 20 GHz bandwidth arbitrary waveform generator (AWG) with a 64 GSa/s sampling rate, then upconverted to the optical domain via an IQ modulator with a 25 GHz bandwidth. A 1550 nm wavelength laser serves as the light source. An optical booster amplifier operates in constant power mode at the IQ modulator output. The ABC device generates dither signals at 20 kHz and 60 kHz. After 80 km of transmission, a variable optical attenuator (VOA) at the receiver (Rx) controls the received optical power (ROP). We use a pre-amplified receiver consisting of an Erbium-Doped Fiber Amplifier (EDFA), an optical bandpass filter, and a photodiode without a trans-impedance amplifier to detect the optical signal. The electrical waveform is captured by a real-time digital storage oscilloscope (DSO) operating at 80 GSa/s.

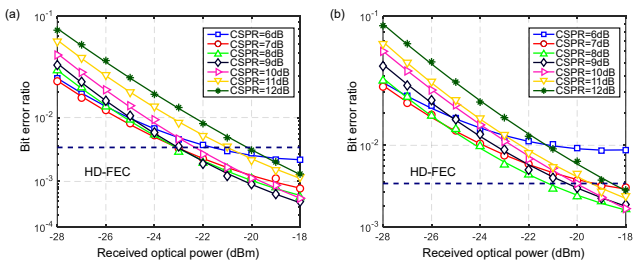

Fig. 5. Measured receiver sensitivity (@BER = 3.8×10⁻³) versus CSPR under (a) back-to-back condition, (b) 80-km transmission condition.

We identified the optimal CSPR for both BtB and 80-km transmission setups, as shown in Fig. 5. First, we calculated the CSPR using the variance and mean of the transmit signal [11]. The system reached optimal performance at a CSPR of 9 dB under BtB conditions; however, the optimal performance shifted to 8 dB after 80 km of standard single-mode fiber (SSFM) transmission. As a result, the experiment used fixed CSPR values of 9 dB for BtB and 8 dB for the 80-km fiber transmission link.

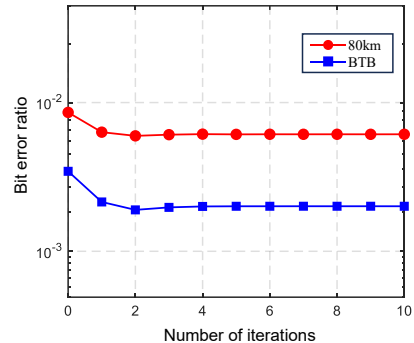

Fig. 6. Measured BER as a function of the iteration number in the mismatch noise cancellation.

We then investigate the performance of DSBIC as a function of the number of iterations. Fig. 6 illustrates the iterative effect of the algorithm's cancellation, with the amplitude of the dither signal set to 5% $V_\pi$ and ROP at -21 dBm for both BtB and 80-km transmission. It is clear that each iteration improves the system's performance, and no further reduction in BER was observed beyond the third iteration. Therefore, we fixed the number of iterations at 3 in subsequent experiments. It can be attributed to the KK's ability to eliminate the nonlinear SSBI distortions. A similar conclusion can be found in [12].

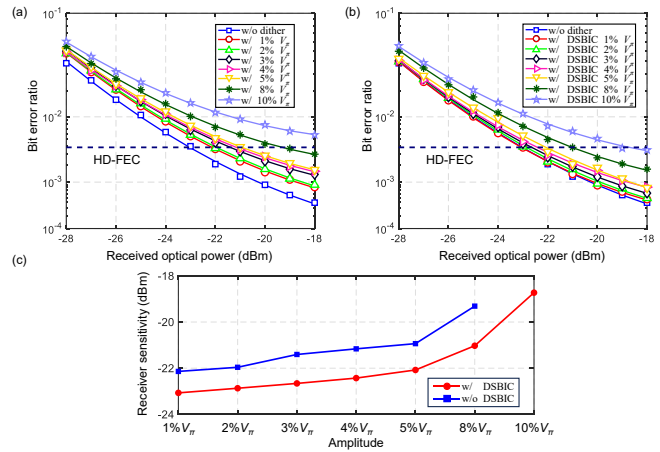

Fig. 7. BER versus amplitude of dither signal at different ROPs in the BtB case using (a) conventional KK receiver, (b) proposed DSBIC algorithm, (c) measured receiver sensitivity (@BER = 3.8×10⁻³) versus different amplitudes of dither signal for w/ and w/o DSBIC algorithm.

We measure the BER curve with different dither amplitudes for the BtB case. As shown in Fig. 7(a) and 7(b), the system transmission performance is highly consistent with the performance degradation caused by the increase in dither amplitude. According to the experimental results, the transmission system can meet the hard-decision feedforward error correction (HD-FEC) threshold ($3.8\times10^{-3}$) even when the dither amplitude reaches 8% $V_\pi$. However, at higher dither amplitudes, a significant degradation in the system's transmission performance becomes apparent. After applying the DSBIC algorithm, we can increase the system's tolerance for dither amplitude to 10% $V_\pi$. When the dither amplitude is low (e.g., 1% $V_\pi$), the system can be restored to its ideal BtB state. Fig. 7(c) shows the system's receiving sensitivity, indicating that the DSBIC algorithm can improve the system's receiving sensitivity by more than 1 dB.

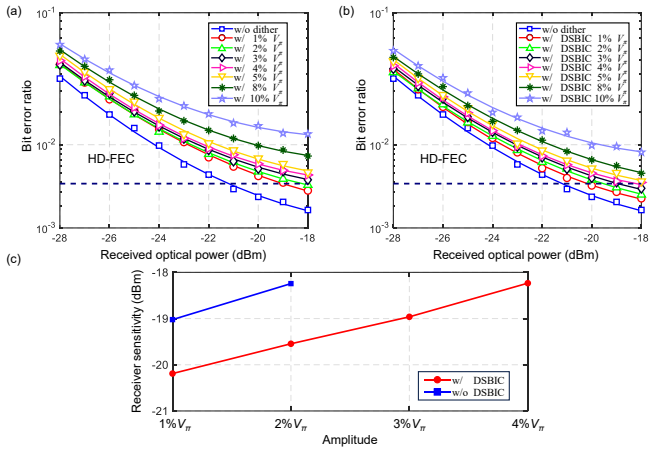

Fig. 8. BER versus amplitude of dither signal at different ROPs in the 80-km transmission using (a) conventional KK receiver, (b) proposed DSBIC algorithm, (c) measured receiver sensitivity (@BER = $3.8\times10^{-3}$) versus different amplitudes of dither signal for w/ and w/o DSBIC algorithm.

Fig. 8(a) and 8(b) display the BER curve with different dither amplitudes for the 80-km transmission case. Compared to the BtB condition, the transmission system can only tolerate a dither amplitude less than 2% $V_\pi$ over an 80-km fiber link. This is because the high-amplitude dither signal used in the ABC equipment can easily excite nonlinear effects during optical fiber transmission, resulting in degraded transmission performance. And the lower dither amplitude has a significant impact on the control stability of the ABC device. According to our previous research [8], dither amplitudes above 2% $V_\pi$ are necessary for stable control with the ABC device. After implementing the DSBIC algorithm, we can increase the system's tolerance for dither amplitude to 4% $V_\pi$, greatly enhancing the control accuracy of the ABC device. Additionally, Fig. 8(c) shows the receiving sensitivity of the 80-km transmission system. In the 80 km fiber transmission scenario, the DSBIC algorithm can improve the system's receiving sensitivity by more than 1 dB, similar to the BtB scenario.

From another perspective, we observed that when the dither was at a lower amplitude, processing the received signal using the proposed cancellation scheme resulted in improved transmission performance with fewer pilot signals. Using the pilot frame decoding method, we achieved better transmission performance with fewer pilot signals. For instance, when the dither amplitude was set to 1% $V_\pi$, the system's receiving sensitivity improved by approximately 1 dB. This is because the algorithm reduced the fluctuations caused by the dither signal, and the concentrated channel estimation suppressed the system noise, resulting in better equalization performance.

In conclusion, we have proposed and experimentally validated a novel iteration noise cancellation method to reduce interference in SSB transmission using commercial ABC modules. This enhances the receiver sensitivity of the system and allows for higher dither amplitudes that are not possible without DSBIC processing. Tests with various dither signal amplitudes confirmed the effectiveness of the proposed DSBIC algorithm, showing an increase in the system's tolerance to dither amplitudes of up to 4% $V_\pi$ in 80-km SSMF transmission.

**Funding.** National Key Research and Development Program of China (2024YFB2908100); National Natural Science Foundation of China (62101049, 62105032); Open Research Fund of Key Laboratory of All Optical Network and Advanced Telecommunication Network, Ministry of Education, Beijing Jiaotong University (AON2024K03).

**Disclosures.** The authors declare no conflicts of interest.

**Data availability.** Data underlying the results presented in this paper are not publicly available at this time but may be obtained from the authors upon reasonable request.